\documentclass[journal,twocolumn]{IEEEtran}
\usepackage{epsfig,makeidx,color}
\usepackage{amsmath,amssymb,bbm}
\usepackage{cite,graphicx,lipsum}
\usepackage{enumerate}
\usepackage[switch,pagewise]{lineno}
\usepackage{hyperref}
\usepackage[ruled,vlined]{algorithm2e}

\usepackage{subfigure,float,booktabs,xfrac}
\newcommand{\jh}[1]{{\textcolor{cyan}{#1}}}

\hypersetup{
        colorlinks = true,
        citecolor=blue,
}
\newcommand{\indicator}{\mathbbm{1}}

\def\uZ{{\mathbb Z}}

\def\uE{{\mathbb E}}

\DeclareMathOperator*{\argmin}{\arg\!\min}

\newtheorem{mylemma}{\bf Lemma} 
\newtheorem{myproposition}{\bf Proposition} 
\newtheorem{corollary}{\bf Corollary} 

\def\deft{ \buildrel \triangle \over = }

\def\be{ \begin{equation} }
\def\ee{ \end{equation} }
\def\bea{ \begin{eqnarray} }
\def\eea{ \end{eqnarray} }

\def\b0{{\bf 0}}

\def\cR{{\cal R}}

\def\cN{{\cal N}}

\def\cW{{\cal W}}

\ifCLASSOPTIONonecolumn
  \newcommand{\figwidth}{0.50\columnwidth}
  \newcommand{\tabwidth}{0.55\columnwidth}
\else
  \newcommand{\figwidth}{0.85\columnwidth}
  \newcommand{\tabwidth}{0.9\columnwidth}
  
\fi

\begin{document}

\title{Explore-Before-Talk:
Multichannel Selection Diversity 
for Uplink Transmissions in Machine-Type Communication}

\author{Jinho Choi, Jihong Park, and Shiva Pokhrel\\
\thanks{The authors are with
the School of Information Technology,
Deakin University, Geelong, VIC 3220, Australia
(e-mail: {jinho.choi, jihong.park, shiva.pohkrel}@deakin.edu.au).
This research was supported 
by the Australian Government through the Australian Research 
Council's Discovery Projects funding scheme (DP200100391).}}

\maketitle
\begin{abstract}
Improving the data rate of machine-type communication (MTC) 
is essential in supporting emerging Internet of things (IoT) applications 
ranging from real-time surveillance to edge machine learning. 
To this end, in this paper we propose a 
resource allocation approach for uplink transmissions
within a random access
procedure in MTC by exploiting multichannel selection diversity, 
coined \emph{explore-before-talk (EBT)}. 
Each user in EBT first sends pilot signals through
multiple channels that are initially allocated 
by a base station (BS) for exploration,
and then the BS informs a subset of initially allocated channels
that are associated with high signal-to-noise ratios (SNRs) 
for data packet transmission by 
the user while releasing the rest of the channels
for other users. 
Consequently, EBT exploits a multichannel selection diversity gain 
during data packet transmission, at the cost of exploration 
during pilot transmission. We optimize 
this exploration-exploitation trade-off, 
by deriving closed-form mean data rate and resource 
outage probability expressions. Numerical results corroborate 
that EBT achieves a higher mean data rate while satisfying 
the same outage constraint, compared 
to a conventional MTC protocol without exploration.
\end{abstract}


\ifCLASSOPTIONonecolumn
\baselineskip 28pt
\fi

\section{Introduction}
The Internet of things (IoT) has sparked 
the upsurge in uplink machine-type communication (MTC) 
traffic~\cite{Ding_20Access} \cite{Al19}. 
The conventional wireless systems postulate their sporadic 
traffic with small payload size (e.g., thermometers, barometers) \cite{3GPP_MTC}. However, emerging IoT applications are envisaged to require higher MTC transmission rates. For instance, e-health wearables generate time-series data. 
Real-time surveillance cameras report video frames. 
Furthermore, edge learning devices exchange on-device 
machine learning models, some of which comprise millions 
of model parameters~\cite{Park:2018aa}. 

\begin{figure}
        \centering
        \subfigure[Conventional ($W=\bar{W}$).]{\includegraphics[width=\columnwidth]{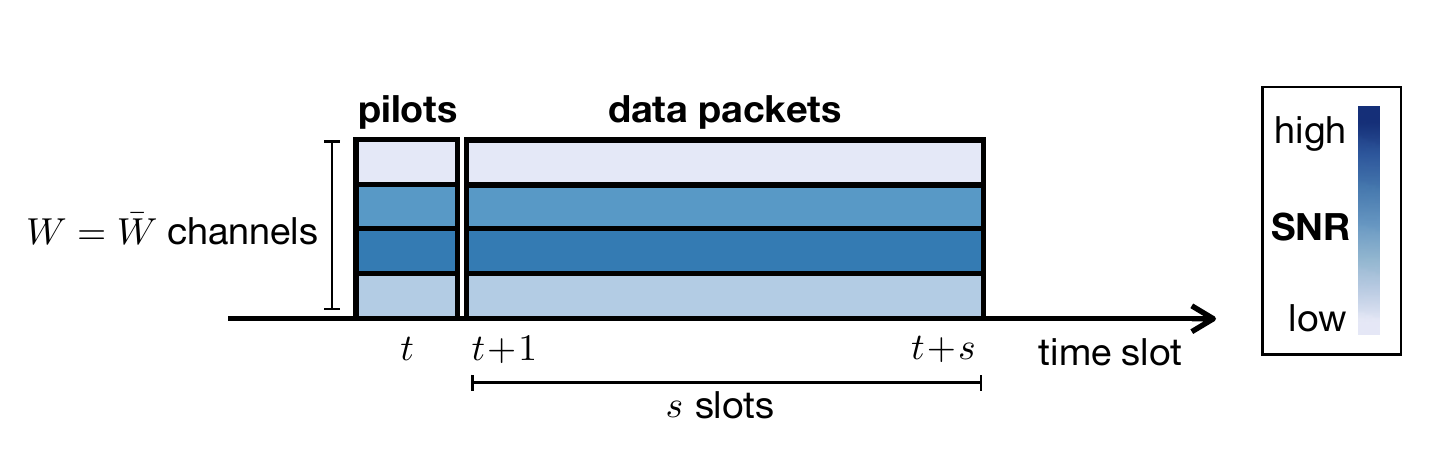}} \vskip -3pt 
        \subfigure[EBT ($W>\bar{W}$).]{\includegraphics[width=\columnwidth]{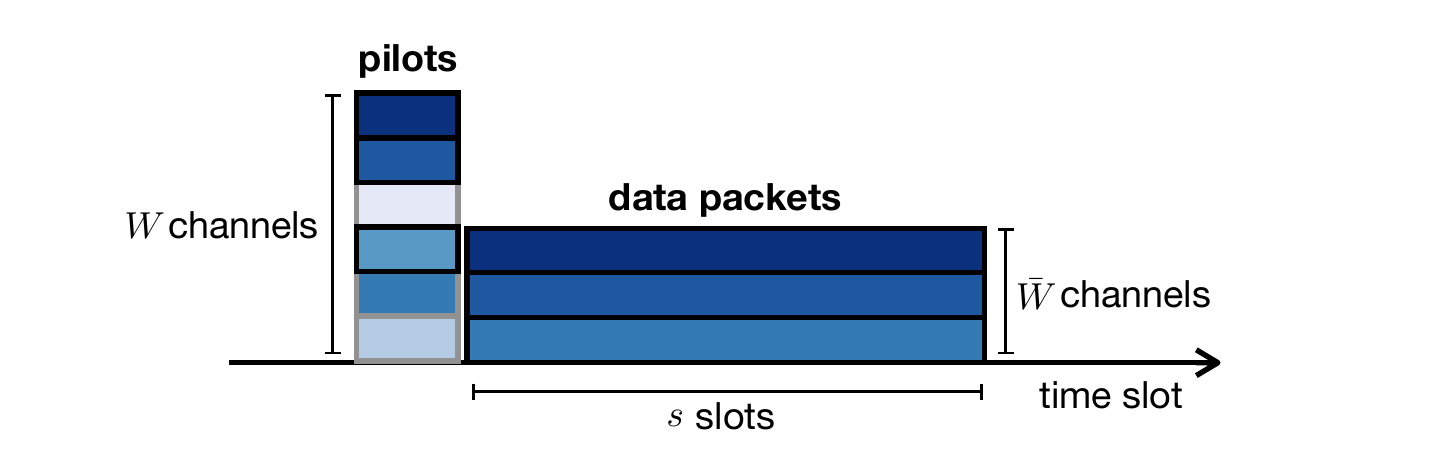}}
        \caption{An illustration of (b) \emph{explore-before-talk (EBT)} exploiting a multichannel selection diversity gain during $s$ slots, obtained from the exploration at slot $t$, compared to (a) a conventional MTC uplink allocation without exploration.}
                \label{Fig:Fig1}
\end{figure}

In MTC, random access approaches are considered to support
a number of MTC devices. In particular, 4-step random access procedure 
is considered in \cite{3GPP_MTC_18}.
In the first step, devices or  MTC user equipment (UE) 
is to randomly choose 
a preamble within a pool of preambles (that is to be shared with
all UEs). After the response from a base station (BS) in the
second step, the UEs can transmit their data packets
through allocated channels to the BS in the third step.
In this 
paper we propose a resource allocation approach for uplink transmissions
in the third step, termed \emph{explore-before-talk (EBT)}. 
As illustrated in Fig.~\ref{Fig:Fig1} (b), 
at the first phase of EBT, every MTC UE 
sends pilot signals through $W (\ge 1)$ channels that are initially allocated
by the BS for exploration,
and then the BS informs a subset of initially allocated channels
that are associated with high signal-to-noise ratios (SNRs),
say $\bar W$ channels of highest SNRs, where $\bar W \le W$,
for data packet transmission by the UE 
while releasing the rest of the channels
for other UEs.
Then, at the second data packet transmission phase, each UE 
uploads data only using $\bar{W}$ highest SNR channels.

Compared to a conventional approach whose $W=\bar{W}$ (i.e., no exploration) as depicted in Fig.~\ref{Fig:Fig1}(a), EBT exploits a \emph{multichannel selection diversity} gain during the data transmission phase, 
at the cost of the channel exploration during the 
pilot transmission phase. To illustrate this, consider an example with three unit-bandwidth channels, in which one good channel provides the unity spectral efficiency, while the other two poor channels have 
ignorable spectral efficiencies. 
In the conventional approach with $\bar{W}=W=1$, a single channel is uniformly randomly selected, and therefore the mean data rate is $\sfrac{1}{3}$. By contrast, in EBT with $W=3$ and $\bar{W}=1$, the mean data rate achieves $1$ by selecting the best channel after exportation. With less exploration, $W=2$, the mean data rate decreases to $\sfrac{2}{3}$, showing the trade-off between the exploring bandwidth and the data rate at exploitation.

Focusing on this exploration-exploitation trade-off, we aim to answer the question whether EBT achieves higher data rate, by fairly comparing the conventional method and EBT under a common traffic model and a resource outage requirement.

\vspace{3pt}\noindent \textbf{Related Works}.\quad
To improve the uplink performance of MTC, random access techniques have been 
studied, such as device grouping methods \cite{Kwon12}, 
coded random access \cite{Paolini15}, 
non-orthogonal multiple access (NOMA) \cite{Choi_JSAC}, and 
sparse code multiple access (SCMA) \cite{Bockelmann16}. 
While the variety of works have addressed reducing random access collision, 
improving MTC data rates has recently attracted much attention 
from both academia and industry, towards supporting emerging applications 
categorized as broadband IoT \cite{Ericsson:19}. 
Next, in terms of the exploring operations before exploitation, 
this work has been partly inspired by listen-before-talk (LBT) 
in dynamic spectrum access (DSA), in which secondary users 
sense the channel before transmission, in order to avoid 
the channels occupied by primary users~\cite{Song:16}. 
A resource allocation method for MTC 
is discussed in \cite{Rekhissa19} 
to improve the energy efficiency by taking into
account the SNR or channel quality indicator (CQI).
In our proposed EBT, MTC users explore the channels providing 
the highest SNRs for data transmission, which is
similar to that in  \cite{Rekhissa19}.
However, we consider multiple channels per MTC user
to exploit selection diversity as illustrated in Fig.~\ref{Fig:Fig1}
(while in \cite{Rekhissa19}, one channel per MTC user
is considered).
Lastly, ordered selection rules have been applied 
in various applications. 
In \cite{Seo:NOMA19} channel inversion power control 
is performed in the order of SNR to improve energy efficiency in uplink NOMA. 
In \cite{Chen:MIMO19}, interfering signals 
are canceled in order of their eigenvalues, thereby efficiently 
utilizing the multi-antenna degrees of freedom. 
In EBT, data transmission channels are selected 
in order of SNR for maximizing the data rates under limited bandwidth.
 
It is noteworthy that the proposed approach
is suitable for MTC UEs with a number of data packets for uplink transmissions,
e.g., data packets to exchange machine-learning models \cite{Park:2018aa}.
On the other hand, there would be a large
number of MTC UEs that have short messages.
For them, 2-step random access approaches are 
preferable  \cite{Choi17IoT} \cite{Abebe17} \cite{Bockelmann18}
\cite{Choi20b}.

\vspace{3pt}\noindent \textbf{Contributions and Organization}.\quad
The contributions of this paper
can be summarized as follows.
\begin{itemize}
\item We propose a resource allocation approach based on
multichannel selection diversity in MTC, i.e., EBT, and derive its data rate 
in closed-form for given $W$ and $\bar{W}$ (see \textbf{Proposition 1})
to see the performance gain by selection diversity.

\item We provide the optimality condition 
of $\bar{W}$ guaranteeing a target outage probability in an overall system
is provided (see \textbf{Proposition~2}), thereby proposing an algorithm 
that finds the optimal $\bar{W}$ for a given $W$ (see \textbf{Algorithm 1}).

\item Numerical evaluations corroborate that EBT achieves higher mean transmission rate with less bandwidth while abiding by the same outage constraint compared to the conventional method.
\end{itemize}

The rest of the paper is organized as follows. In Section II, the system model for EBT is presented. In Section III, the mean data rate is studied. In Section IV, the exploration and exploitation (i.e., $W$ and $\bar{W}$) are optimized for a given outage requirement, which is validated by numerical evaluations in Section IV, followed by our conclusion in Section V.

\section{System Model}

In this section, we present the system model to study
resource allocation for uplink in MTC with multiple 
UEs and a BS.

\begin{figure}
\begin{center}
\includegraphics[width=\figwidth]{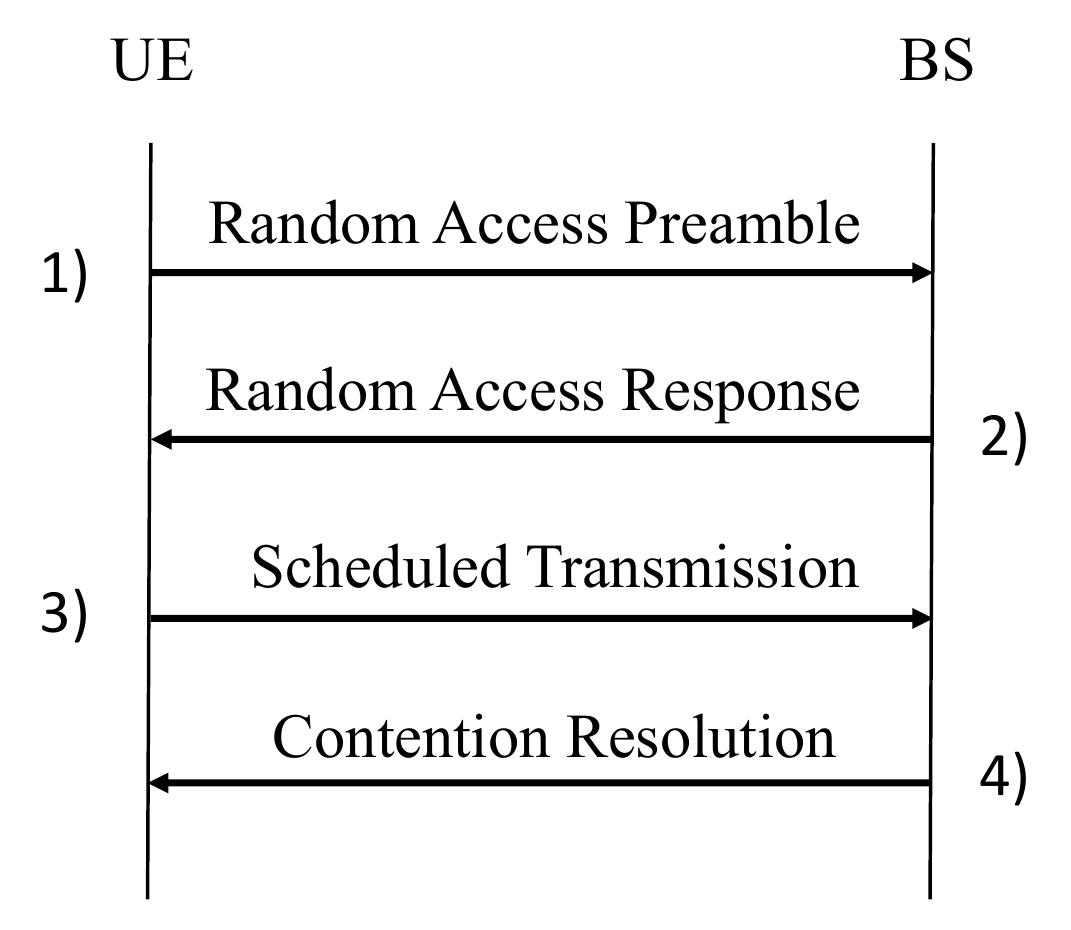}
\end{center}
\caption{4-step random access procedure in MTC.}
                \label{Fig:4step}
\end{figure}

Suppose that uplink transmissions are carried out 
based on 4-step random access procedure \cite{3GPP_MTC} \cite{3GPP_NBIoT},
which is illustrated in Fig.~\ref{Fig:4step}.
The first step is random access 
to establish connection to the BS
with a pool of preambles consisting of $L$ sequences.
In the first step, an active UE 
transmits a randomly selected a preamble 
through physical random access channel (PRACH).
In the second step, the BS detects the transmitted
preambles and sends responses. Once 
an active UE is connected to the BS, it can transmit data packets
in the third step
through dedicated resource blocks (RBs)
or channels\footnote{We will use 
the terms resource block and channel interchangeably.},
which are physical uplink shared channel (PUSCH).

\begin{figure}
\begin{center}
\includegraphics[width=\figwidth]{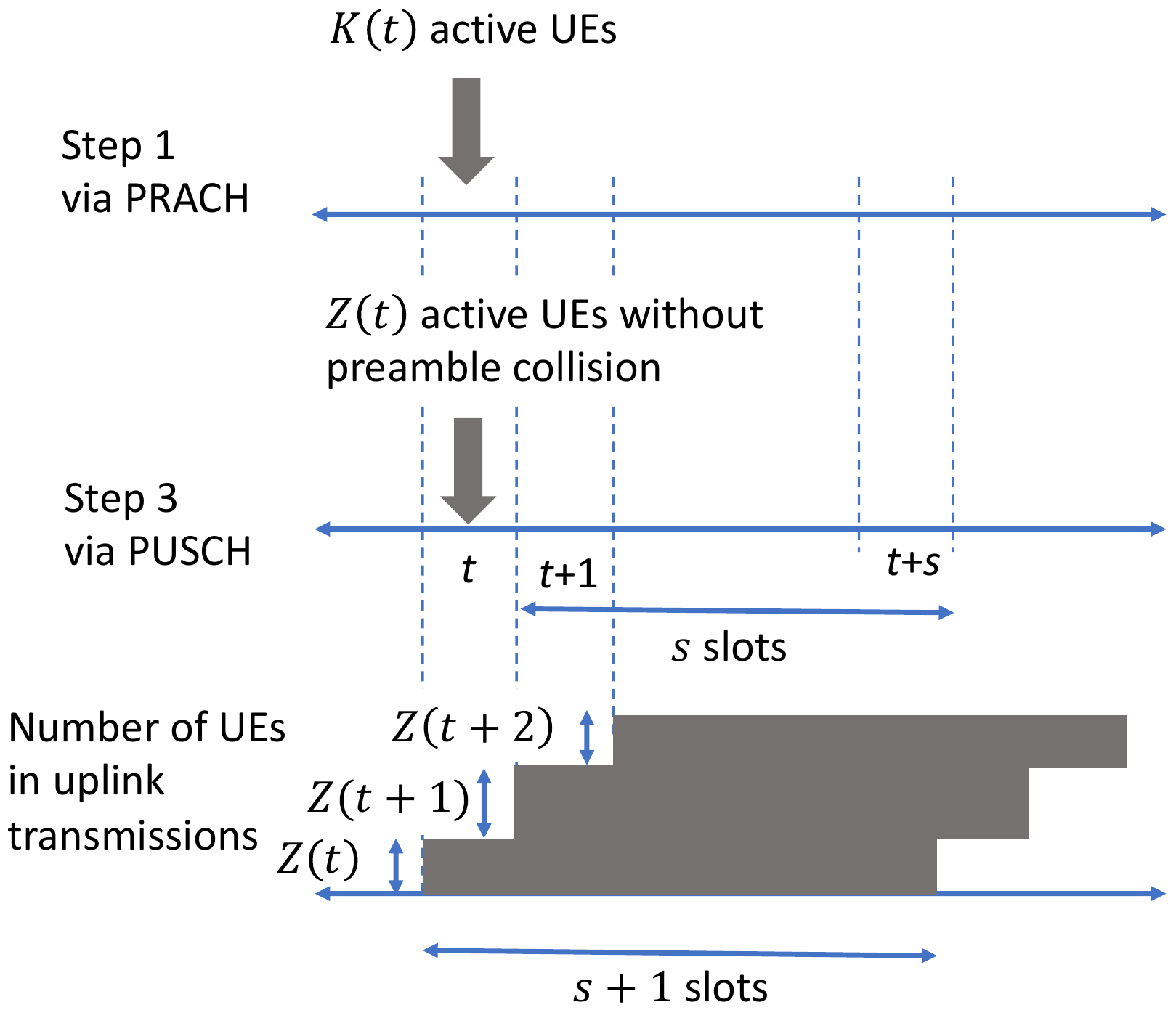}
\end{center}
\caption{4-step random access procedure in MTC.}
                \label{Fig:Flow}
\end{figure}

In this paper, we focus on the third step 
with preamble detection carried out in the first step
as in \cite{Jang17}. Thus, it is assumed that
the UEs without preamble collision can move to the third step,
while the UEs with preamble collision
would be backlogged.
Note that in this case, the fourth step can be absorbed into the
second step.
As shown in Fig.~\ref{Fig:Flow}, at slot $t$, there are $K (t)$
active UEs that transmit preambles. 
Among the $K(t)$ UEs in the first step, 
$Z(t)$ UEs are admitted for uplink transmissions in the third step. 
Here, $Z(t) \le K(t)$ due to preamble collisions.
Thus, every time slot $t$, it is necessary to allocate
RBs to $Z(t)$ newly admitted UEs (without preamble collision)
for uplink transmissions (in the third step).
It is assumed that each newly admitted UEs
needs $s+1$ slots for uplink transmissions.
As will be explained later, the first slot is used to
transmit a pilot signal to allow the BS to estimate
the UE's channel in PUSCH.
In addition, $s$ can vary from one UE to another
as will be considered in Section~\ref{S:EE}.

Suppose that there are $N$ radio RBs for 
the UEs in the third step, i.e., uplink transmissions.
Let $h_{n,k}$ denote the channel coefficient of the $n$th 
channel from UE $k$ to the BS.
Note that the newly admitted UEs in the third step
transmit their signals through PUSCH, which 
is different from PRACH. 
Thus, the channel state information (CSI) estimation 
in the first step with preamble transmission
does not provide the estimates of the $h_{n,k}$'s (which
are the channel coefficients the RBs of PUSCH).
This means that a newly admitted UE
in the third step needs to 
transmit pilot signals
together with data packets
over $s$ slots
as illustrated in Fig.~\ref{Fig:Fig1} to allow the BS
to estimate $h_{n,k}$.
Thus, for each UE in the third step,
a total of $1+s$ slots are required.
For convenience, let $\gamma_{n,k} = \frac{|h_{n,k}|^2}{\sigma^2}$
be the signal-to-noise ratio
(SNR) of the $n$th channel from UE $k$ to the BS,
where $\sigma^2$ is the noise variance.


With known CSI of all the UEs
in the third step, the BS can perform optimal channel allocation with $N$ RBs.
For example, suppose that $N = 1000$ and each UE needs to 
have $10$ RBs for uplink transmissions.  Then,
there can be up to $100$ UEs that are connected simultaneously.
In this case, in order to allocate $10$ RBs for each UE,
the CSI estimation of all $N = 1000$ channels
per UE is required, which may require excessive pilot signaling.
In addition, since the optimal channel
allocation has to be updated when a set of new UEs join in every slot,
it requires not only
a high computational complexity,
but also existing UEs to change their channels for uploading
in every slot.


To avoid the difficulties stated above,
a random subset of available channels can be simply allocated
to a new admitted UE for uplink transmissions
in the third step, which is regarded as a conventional approach.

\section{Multichannel Selection Diversity Gain in Mean Data Rate}	
\label{S:MSD}

In this section, we mainly focus on the performance
gain of each UE in the third step by EBT, which can exploit multichannel
selection diversity, in terms of the mean rate 
for given $W$, $\bar W$, and $s$ as in Fig.~\ref{Fig:Fig1}.
To understand the overall performance,
it is also necessary to find the number of UEs
in the third step for given $N$, which will be studied
in Section~\ref{S:EE}.


Suppose that the BS randomly chooses
$W$ channels and allocates them to a new UE, say UE $k$.
For convenience, let $\cW = \{1, \ldots, W\}$
be the index set of the allocated $W$ channels.
Note that since $\cW$ is a set of randomly selected
channels from $\cN = \{1, \ldots, N\}$ as mentioned earlier, 
$\cW$ can be any subset of $\cN$.
Then, UE $k$ transmits pilot signals through
the $W$ channels so that the BS can estimate their SNRs.
The $\bar W$ best channels among $W$
in terms of SNR are selected and their indices
are fed back to UE $k$ so that the UE can transmit their
data packets through the selected $\bar W$ channels
as shown in Fig.~\ref{Fig:Fig1},
where $s$ represents the number of slots UE $k$
sends packets.
In doing so, EBT can exploit the multichannel
selection diversity gain to increase
the transmission rate, where
$W$ and $\bar W$
are referred to as the number of the channels for exploration 
(to select the best $\bar W$ channels)
and
exploitation (to transmit
data packets), respectively.

In MTC, since UEs may have short messages to transmit and
their mobility is limited, 
throughout the paper,
it is assumed that 
\emph{i)} 
each UE transmits packets over a finite number of slots,
i.e., $s$ is finite;
\emph{ii)} the channel coefficients
remain unchanged for $s$ slots.

For convenience, we omit the index of UE $k$ in the rest of this section.
Let 
\be
\gamma_{(W)} \ge \ldots \ge \gamma_{(\bar W+1)} \ge\gamma_{(\bar W)} \ge
\ldots \ge \gamma_{(1)},
\ee
where $\gamma_{(w)}$ represents the $w$th lowest element.
Then, for given $\{\gamma_w\}$, the $\gamma_{(w)}$'s are 
order statistics \cite{DavidBook}.
Since the UE will keep the channels of 
the $\bar W$ highest SNRs, the mean transmission 
rate (in bits per second per Hz) per slot becomes
\begin{align}
R (W,\bar W) 
& = \uE\left[\sum_{i=W-\bar W+1}^{W} \log_2 (1+ \gamma_{(i)}) \right] \cr
& = 
\sum_{i=W-\bar W +1}^{W} \uE[ \log_2 (1+ \gamma_{(i)}) ].
	\label{EQ:RWW}
\end{align}
In \eqref{EQ:RWW},
we assume that the transmission rate of each
selected channel is set to its achievable rate
(i.e., the channel capacity \cite{CoverBook}).

For comparisons, consider a conventional approach
where $W_{\rm c}$ channels are randomly allocated to a UE
regardless of their SNRs.
This can be seen as a special case of EBT with $W = \bar W = W_{\rm c}$.
Due to the random selection of $W_{\rm c}$ channels, 
it can be assumed that $\gamma_w$, $w \in \cW_{\rm c}
= \{1, \ldots, W_{\rm c}\}$, is independent.
Furthermore, for tractable analysis, we assume
that $\gamma_w$ is independent and identically distributed (iid).
In this case, the mean transmission
rate of the conventional approach becomes
\be
R_{\rm c} =  W_{\rm c} \uE[\log_2 (1+ \gamma_w)].
	\label{EQ:R_c}
\ee

For example, suppose that the UE has the SNR of channel
$w \in \cW$ that is distributed as follows:
\be
f(\gamma_w) = \frac{1}{\bar \gamma} \exp
\left( - \frac{\gamma_w}{\bar \gamma} \right), \ \gamma_w > 0,
	\label{EQ:Ray}
\ee
where $\bar \gamma$ is the mean SNR.
That is, each channel is modeled as an independent and identical 
Rayleigh\footnote{Other fading channels
can be considered for $\gamma_w$,
while we consider Rayleigh fading in this section
as an example to demonstrate the gain from selection
diversity in the third step.
It is also assumed
that power control is used to compensate large-scale fading. 
As a result, all $W$ channels have the same mean SNR.}
fading channel
and $\gamma_w$, $w \in \cW$, is 
seen as a realization from the distribution in \eqref{EQ:Ray}. 
Then, from \cite{Alouini99}, we have
\begin{align}
R_{\rm c} = W_{\rm c} 
\frac{e^{\frac{1}{\bar \gamma}}}{\ln 2} E_1 \left(
\frac{1}{\bar \gamma}
\right),
\end{align}
where $E_n (x) = \int_1^\infty \frac{e^{-x t}}{t^n} d t$ 
is the exponential integral function of order $n$.

In a similar way, we derive the mean transmission rate 
of EBT in closed-form in the following lemma.
\begin{mylemma} \label{L:1}
For independent Rayleigh fading in
\eqref{EQ:Ray}, the mean transmission rate (per slot) of EBT is given as:
\begin{align}
        R(W, \bar W) & =  \sum_{w = W - \bar W + 1}^W
        \frac{W!}{(W-w)!} \cr
        & \!\! \times
        \sum_{m=0}^{w-1} \frac{(-1)^m }{(w-1-m)! m!} 
        \frac{e^{-\frac{a_{m,w}}{\bar \gamma}}}{
        a_{m,w} \ln 2} E_1 \left(\frac{a_{m,w}}{\bar \gamma} \right), \ \ 
                \label{EQ:RWWc}
        \end{align}
        where $a_{m,w} = W - w + 1 + m$.

\end{mylemma}
\begin{IEEEproof}
From \eqref{EQ:RWW}, it can be shown that
\begin{align}
R (W,\bar W) = \sum_{w = W -\bar W +1}^W
\int_0^\infty \log_2 (1+ \bar \gamma x) f_{(w)} (x) dx ,
	\label{EQ:RWWi}
\end{align}
where 
\be
f_{(w)} (x) = \frac{W!}{(w-1)!(W-w)!}
(1 - e^{-x})^{w - 1} e^{- (W-w+1)x},
	\label{EQ:fw}
\ee
which is the probability density function (pdf)
of the $w$-th order statistic \cite{DavidBook}.
Applying the binomial expansion to \eqref{EQ:fw}, we obtain
\begin{align}
f_{(w)} (x) = \frac{W!}{(w-1)!(W-w)!} 
\sum_{m=0}^{w-1} \binom{w-1}{m} 
(-1)^{m} e^{- a_{m,w}x}.
        \label{EQ:fwb}
\end{align}
Plugging this into \eqref{EQ:RWWi} completes the proof.
\end{IEEEproof}

Next, for a large $W$, 
we can further simplify the mean transmission rate of EBT as follows.

\begin{myproposition} \label{L:2}
For independent Rayleigh fading in
\eqref{EQ:Ray} with $W \gg \bar{W} \gg 1$, the mean transmission rate of EBT is:
\begin{align}
        R(W, \bar W)      \approx \sum_{i=1}^{\bar W}  \log_2 \left(1+ \bar \gamma
        \ln \frac{W}{i} \right).
	\label{EQ:RWW_approx}
\end{align}
\end{myproposition}
\begin{IEEEproof}
For $W \gg \bar{W} \gg 1$, the $w$-th highest SNR $\gamma_w$ with 
$w\leq \bar{W}$ is large with a high probability. In this regime, the expectation of the logarithmic function increasing with diminishing returns in \eqref{EQ:RWW} is approximated~as
\begin{align}
        \uE[\log_2(1 + \gamma_w)] \approx \log_2(1 + \uE[\gamma_w]). \label{EQ:GammaApprox}
\end{align}
Next, we focus on $\uE[\gamma_w]$ in \eqref{EQ:GammaApprox}. Let $F(\gamma)$ denote the probability
that $\gamma_w \le \gamma$. This cumulative distribution function
(cdf) of the SNR is given by integrating its pdf in \eqref{EQ:fw}.
For independent Rayleigh fading channels, following \cite{DavidBook}, the cdf is bounded as
\be
\frac{w}{W +1} 
\le F(\uE[\gamma_{w}])\le 
\frac{w}{W}. \label{EQ:CDFbounds}
\ee
Therefore, for $W\gg 1$ we obtain
\begin{align}
        F(\uE[\gamma_{w}]) \approx \frac{w}{W}. \label{EQ:CDFapprox}
\end{align}
Since $F(\gamma)$ is monotonically increasing, we can take the inverse function, yielding
\begin{align} 
        \uE[\gamma_{w}] 
        &\approx F^{-1}\left( \frac{w}{W} \right) \cr
        &= \bar{\gamma} \ln \left(\frac{W}{W - w} \right). \label{EQ:QuantileLB}
\end{align}
The last equality follows from the quantile function $F^{-1} (p)$ of the SNR distribution in \eqref{EQ:Ray} with $p={w}/{W} \in [0,1]$. Applying \eqref{EQ:QuantileLB} with \eqref{EQ:GammaApprox} to \eqref{EQ:RWW} and inverting the order of the summation finalize the proof.
\end{IEEEproof}
In \eqref{EQ:RWW_approx}, considering the summation of the smallest values, i.e., $i=\bar{W}$, yields the following corollary.
\begin{corollary}
For independent Rayleigh fading in
\eqref{EQ:Ray} with $W \gg \bar{W} \gg 1$, the mean transmission rate of EBT is lower bounded as:
\begin{align}
        R(W, \bar W) \geq \bar W \log_2 \left(1
+ \bar \gamma \ln \frac{W}{\bar W} \right). \label{EQ:RWW_q}
\end{align}
\end{corollary}

In \eqref{EQ:RWW_q}, we can clearly see the gain from
the multichannel selection diversity as the approximate
rate increases with $W$ when $\bar W$ is fixed. In particular,
we can see that the SNR term in the logarithm
is effectively increased
by a factor of $\ln ({W}/{\bar W})$.

\begin{figure}
\begin{center}
\includegraphics[width=\figwidth]{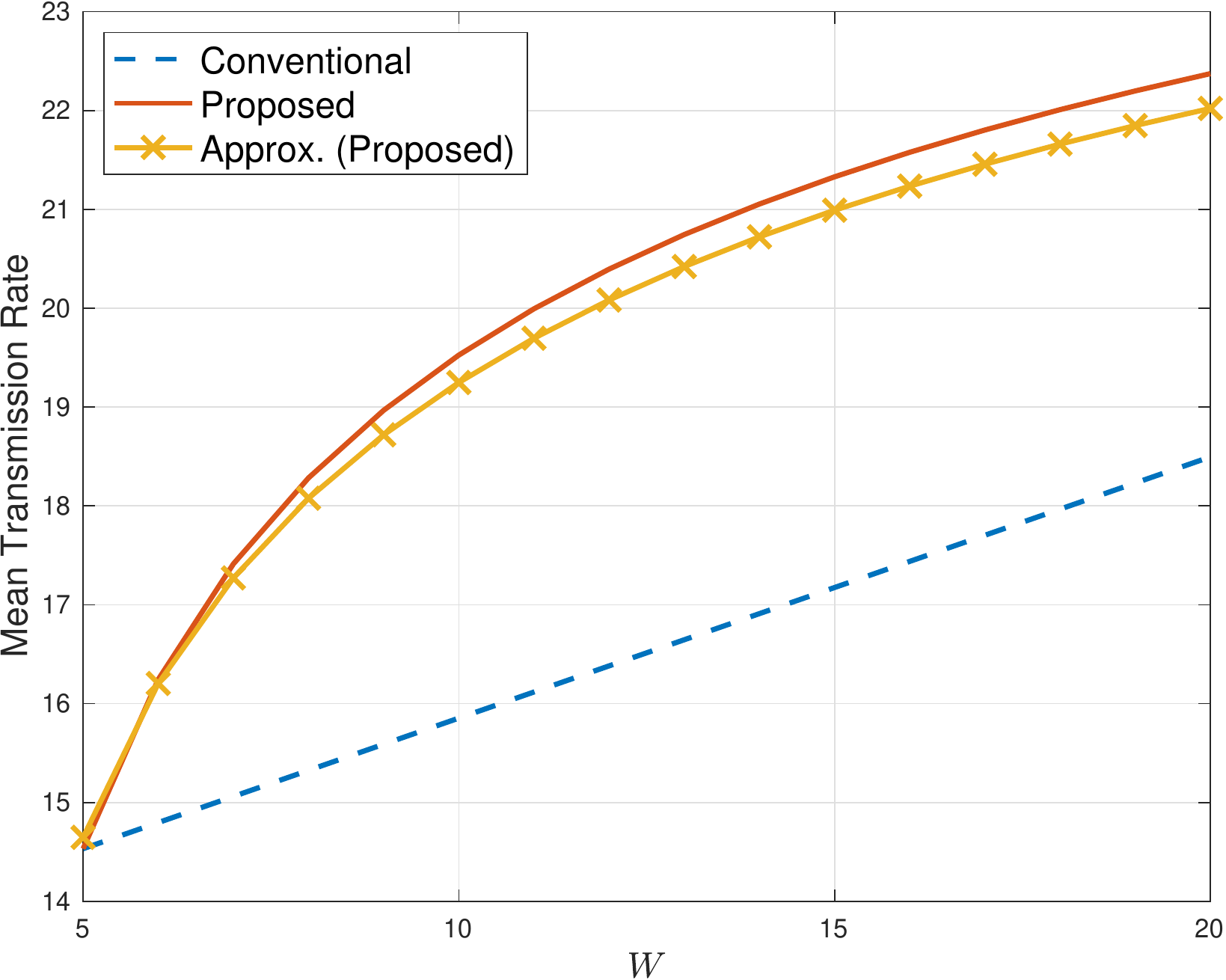}
\end{center}
\caption{Mean transmission rate (per slot) as functions
of $W$ when $\bar W = 5$, $s = 10$, and $\bar \gamma = 10$ dB.}
        \label{Fig:c_rate1}
\end{figure}

For a fair comparison
between the conventional and proposed approaches,
the number of channels of the conventional approach
is decided as
\be
W_{\rm c} = \frac{W + s \bar W}{1+ s}
	\label{EQ:WWs}
\ee
so that the total number of channels 
is the same as that of EBT,
which is $W + s \bar W$.
In Fig.~\ref{Fig:c_rate1},
under the assumption of independent Rayleigh
fading in \eqref{EQ:Ray},
the mean transmission rates (per slot) as functions
of $W$ when $\bar W = 5$, $s = 10$, and $\bar \gamma = 10$ dB
are shown for the conventional and proposed approaches.
Note that in the conventional approach, $W_{\rm c}$
is decided as in \eqref{EQ:WWs} for given $\bar W$ and $W$.
The approximate mean transmission rate of EBT is given by \eqref{EQ:RWW_q} (shown by the solid line with cross markers).
Clearly, for a fixed $\bar W$, 
a higher selection diversity gain can be achieved as $W$ 
increases in EBT,
i.e., the number of channels for exploration
increases, there is a higher chance to choose
the better $\bar W$ channels for exploitation,
which results in a higher mean transmission rate
than that of the conventional one as shown in Fig.~\ref{Fig:c_rate1}.
Note that the mean transmission 
rate of the conventional approach also increases
with $W$ as $W_{\rm c}$ increases with $W$.

\begin{figure}
\begin{center}
\includegraphics[width=\figwidth]{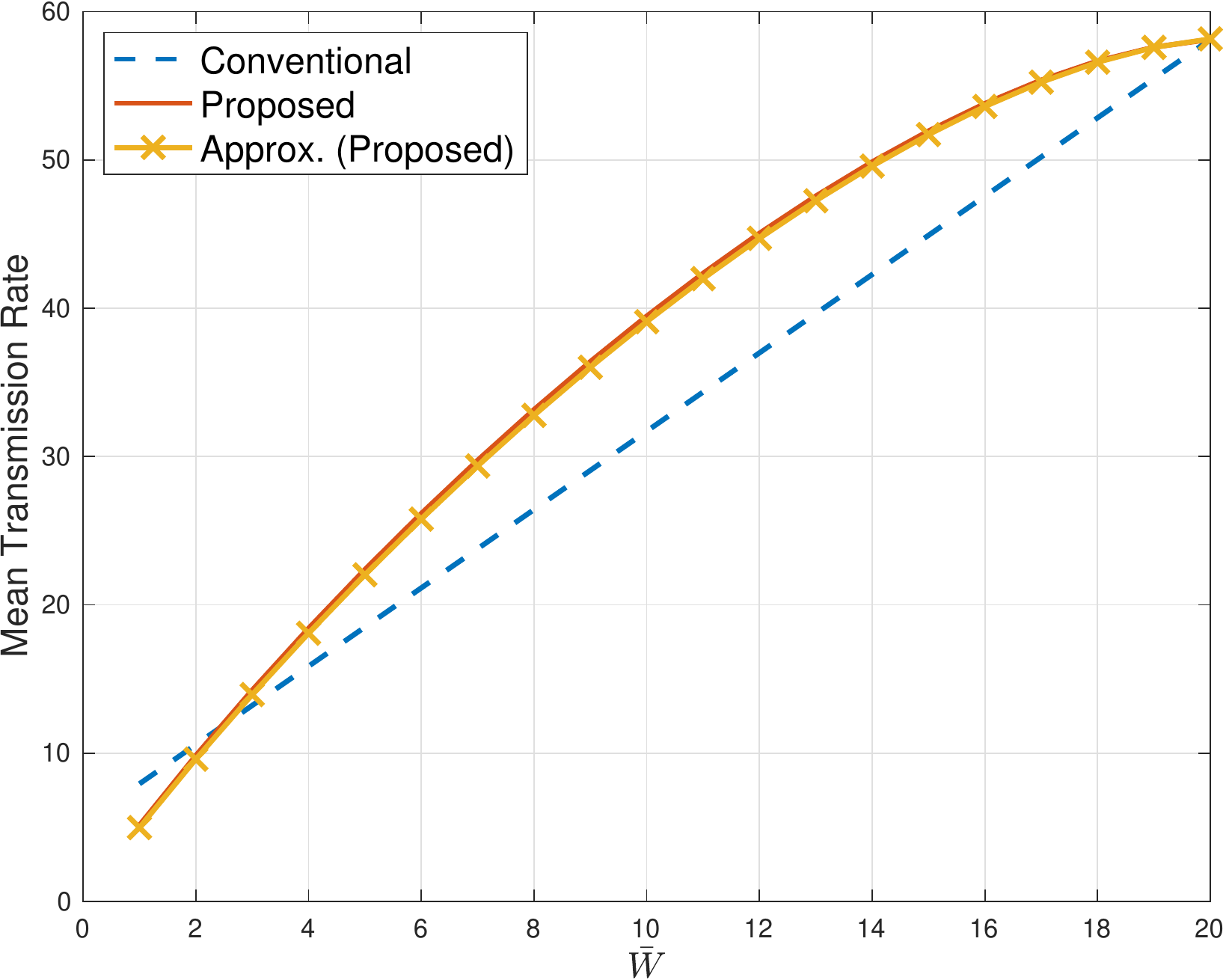}
\end{center}
\caption{Mean transmission rates (per slot) as functions
of $\bar W$ when $W = 20$, $s = 10$, and $\bar \gamma = 10$ dB.}
        \label{Fig:c_rate3}
\end{figure}

In Fig.~\ref{Fig:c_rate3},
the mean transmission rates as functions
of $\bar W$ when $W = 20$, $s = 10$, and $\bar \gamma = 10$ dB
are shown for the conventional and proposed approaches.
It is interesting to see that 
the conventional approach has a better performance
when 
the number of channels for exploitation is small, e.g.,
$\bar W \le 2$. In particular, when $\bar W = 1$
in EBT, $W_{\rm c}$ in the conventional approach becomes 
2.727. Thus, although
the multichannel selection diversity gain can help improve
the transmission rate of EBT,
a large number of channels for exploitation
is still necessary for a higher transmission rate.
It is also noteworthy that
when $W = \bar W = W_{\rm c} = 20$,
EBT becomes
the conventional approach, which results in the same
mean transmission rate.

\section{Exploration-Exploitation Optimization under an Outage Constraint}
\label{S:EE}

As shown in Figs.~\ref{Fig:c_rate1}
and ~\ref{Fig:c_rate3}, if 
the numbers of the channels for exploration and exploitation
(i.e., $W$ and $\bar W$, respectively) are well chosen,
EBT can provide a higher
transmission rate than the conventional approach
subject to~\eqref{EQ:WWs}.
However, from a multiuser system point of view,
for a fair comparison, 
we also need to consider
the number of UEs in the third step for a given number of RBs, $N$.
To this end, in this section,
we consider the overall uplink system that includes
the first and third\footnote{In 4-step random
access procedure, the second and fourth steps are for
downlink, which will be ignored for the
uplink performance analysis.} 
steps, which is illustrated in Fig.~\ref{Fig:Flow}.

The numbers of UEs
in the first and third steps are random variables.
As a result, although a new UE without
preamble collision can move to the third step,
it may not have a sufficient number of channels for uplink
transmissions. This event is referred to as 
the outage event. The system has to be designed
to keep the outage probability low
or the outage probability less than $e^{-d}$, $d > 0$, 
is a design parameter.
Thus, in this section, we first obtain an upper-bound
on the outage probability and then 
consider an optimization problem formulation
for a fair comparison between EBT and the conventional approach
in terms of the mean transmission rate and outage probability.



\subsection{Outage Probability}

This subsection focuses on the outage event at a certain time, when the total $N$ channels are exceeded by the allocated channels $Q$ to newly admitted UEs as well as the remaining UEs who are previously admitted. For mathematical amenability, we consider the following assumptions in the outage probability analysis.
\begin{itemize}
\item[{\bf A1)}] 
Recall that $K(t)$ is the number of new UEs that send preambles
to upload their data to the BS in slot $t$, where
$K(t)$ is Poisson-distributed, i.e.,
\be
K (t) \sim {\rm Pois} (\lambda),
\ee
where $\lambda$ is the mean of the number
of new active UEs per slot. In addition, $K(t)$ is independent for every
$t$.

\item[{\bf A2)}] An admitted UE transmits 
its data over a number of slots and
the number of slots required for each device 
to upload its data varies from one UE to another.
Let $\nu_s$ denote the probability that a 
UE requires $s$ slots to upload its data.
In addition, assume that $s_{\rm max}$ is the maximum
number of slots to upload data from any UE. Thus,
we have
$\sum_{s = 1}^{s_{\rm max}} \nu_s = 1$.
\end{itemize}

Recall that $Z(t)$ is the number of the admitted UEs
(into the 3rd step),
which are the UEs without preamble collisions, 
and $C_l(t)$ the number of active UEs that choose the $l$-th preamble out of $L$ preambles.
Provided that an active UE chooses a preamble uniformly at random,
under {\bf A1},
we can show that
\be
Z(t) = \sum_{l=1}^L \indicator(C_l (t) = 1),
\ee
where 
\be
C_l(t) \sim {\rm Pois} \left(\frac{\lambda}{L} \right).
\ee
Thus, we can see that
$Z(t)$ has the binomial distribution with parameter $L$ and $p$,
denoted by ${\rm Bin}(L,p)$,
or
\be
\Pr(Z(t) = a) = \binom{L}{a} p^a (1 -p)^{L-a},
	\label{EQ:Z_B}
\ee
where
$p = \frac{\lambda}{L} e^{-\frac{\lambda}{L}}$.
The Poisson approximation can be used for $Z(t)$, which
results in the following distribution for $Z(t)$:
\be
Z(t) \sim {\rm Pois} \left(\lambda e^{-\frac{\lambda}{L}} \right).
	\label{EQ:P_Z}
\ee

\begin{figure}
\begin{center}
\includegraphics[width=\figwidth]{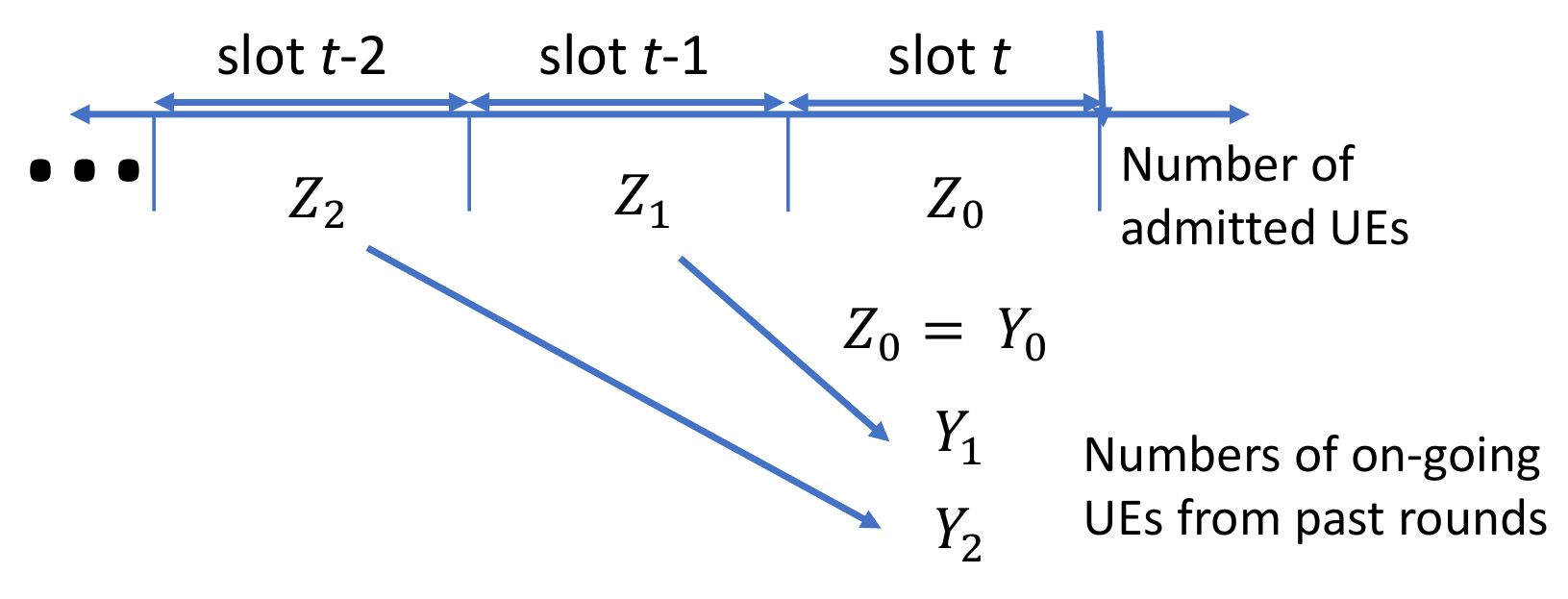}
\end{center}
\caption{The relationship between $Z_s$ and $Y_s$.}
        \label{Fig:Fig2}
\end{figure}

For convenience, let $Z_0 = Z(t)$, and at time slot $t$,
let $Z_s$ be the number of the UEs that are admitted
in the last $s$ slot. Then, $Z_s$ has the same distribution of $Z_0$. In addition, let $Y_s$ be the number of the UEs
that still upload data among $Z_s$. Fig.~\ref{Fig:Fig2} depicts the relationship between $Z_s$ and $Y_s$. It is clear that $Y_0 = Z_0$ 
and $Y_s$ might decrease with $s$ (as more UEs likely 
finish their uploading as $s$ increases). For $s\geq 1$, under {\bf A2} with the Poisson approximation in \eqref{EQ:P_Z}, a fraction $\sum_{i=s}^{s_\text{max}}\nu_i$ of $Z_s$ remains at time $t$, leading to
\be
Y_s \sim {\rm Pois} (\lambda_s),  \ s = 1, \ldots, s_{\rm max},
	\label{EQ:Y_s}
\ee
where $\lambda_s = \lambda e^{-\frac{\lambda}{L}} \sum_{ i=s }^{s_{\rm max}} \nu_i$. 


Finally, using $Y_s$, we can express the number of allocated 
channels at a certain time slot as follows:
\begin{align}
Q & = W Y_0 + \sum_{s = 1}^{ s_{\rm max} } \bar W Y_s \cr
& = W Y_0 + \bar W V,
\end{align}
where $V = \sum_{s = 1}^{ s_{\rm max} }  Y_s$.
Then, 
using the Chernoff bound \cite{Mitz05},
an upper-bound on
the outage probability becomes
\begin{align}
\Pr(Q > N) 
& \le 
 \uE[e^{\theta (Q-N)}] \cr
& = 
 e^{- \theta N} \uE[e^{\theta (W Y_0 + \bar W V)}] , \ \theta \ge 0.
	\label{EQ:CB}
\end{align}
It can also be shown that
\be
\log \Pr(Q > N)  \le \psi (\theta; W, \bar W),
\ee
where 
\begin{align}
        \psi (\theta; W, \bar W) = M_y (W \theta) + M_v (\bar W \theta) - \theta N.
\end{align}
The terms $M_y(\theta) = \log \uE[e^{\theta Y_0}]$
and $M_v(\theta) = \log \uE[e^{\theta V}]$
are the 
logarithmic moment generating functions (LMGFs)
of $Y_0$ and~$V$, respectively. An upper bound on the outage probability is given by
\begin{align}
\Pr(Q > N) 
& \le \min_{\theta \ge 0} 
\exp \left( \psi (\theta; W, \bar W) \right) \cr
& = \exp \left( \min_{\theta \ge 0} 
\psi (\theta; W, \bar W) \right).
	\label{EQ:upp}
\end{align}
It is well-known that the LMGF is convex (e.g., \cite{Kelly_Yudovina}).
Thus, there exists a unique solution to the minimization
on the right-hand side (RHS) in \eqref{EQ:upp}.
We can also have the following monotonic property of $\psi(\cdot)$.

\begin{mylemma}	\label{L:3}
For $\bar W^\prime < \bar W^{\prime \prime} \le W$, it satisfies
\begin{align}
\min_{\theta \ge 0}
\psi (\theta; W, \bar W^\prime ) 
\le
\min_{\theta \ge 0}
\psi (\theta; W, \bar W^{\prime \prime}).
	\label{EQ:ineq}
\end{align}
\end{mylemma}
\begin{IEEEproof}
Let $\psi_1(\theta) = \psi(\theta; W, \bar W^\prime)$ and
$\psi_2(\theta) = \psi(\theta; W, \bar W^{\prime \prime})$.
Since $\bar W^\prime < \bar W^{\prime \prime}$,
we have
$$
\psi_2 (\theta) - \psi_1 (\theta)
= M_v (\bar W^{\prime \prime} \theta) - 
M_v (\bar W^{\prime} \theta) 
= \ln \frac{\uE[e^{\bar W^{\prime \prime} \theta V}]}
{\uE[e^{\bar W^{\prime} \theta V}]} \ge 0
$$
or $\psi_1 (\theta) \le \psi_2 (\theta)$, $\theta \ge 0$.
Let $\theta^* = \argmin_{\theta \ge 0} g_2 (\theta)$.
Note that since the LMGF is convex, $\theta^*$ exists.
Then, we have
\be
\min_{\theta \ge 0} \psi_2 (\theta) = 
\psi_2 (\theta^*) \ge 
\psi_1 (\theta^*) \ge \min_{\theta \ge 0} \psi_1 (\theta),
\ee
which completes the proof.
\end{IEEEproof}

It is noteworthy that if $M_v (\theta)$ is a monotonic increasing
function of $\theta$, we have the strict inequality
in \eqref{EQ:ineq}, i.e., $\le$ is replaced with $<$.
In addition, an upper-bound on
the outage probability of the conventional
approach can be found using
\eqref{EQ:upp}
by letting $W = \bar W  = W_{\rm c}$ as a special case.

\subsection{Optimization and Trade-off}

In this subsection, we consider an optimization problem
to decide 
the numbers of channels for admitted UEs
in the 3rd step to upload their data packets with a QoS constraint.

From \eqref{EQ:upp},
it can be shown that
\begin{align}
\min_{\theta \ge 0} \psi(\theta; W, \bar W) \le - d 
\Rightarrow \Pr(Q > N) \le e^{-d},
	\label{EQ:mint}
\end{align}
where $d > 0$ can be seen as the negative
quality-of-service (QoS) exponent \cite{Chang95} \cite{Kelly_Yudovina}.
For the inequality on the
left-hand side (LHS) term in \eqref{EQ:mint},
we need the following result.

\begin{mylemma}	\label{L:4}
Under the Poisson approximation
in \eqref{EQ:Y_s}, we have
\be
V \sim {\rm Pois} (\lambda e^{-\frac{\lambda}{L}} \bar s),
	\label{EQ:VP}
\ee
where $\bar s = \sum_{s=1}^{ s_{\rm max}} s \nu_s$
is the mean of the number of slots for uploading.
\end{mylemma}
\begin{IEEEproof}
Since the sum of independent Poisson random variables
is also Poisson, we have \eqref{EQ:VP}.
We omit the detail as the derivation of \eqref{EQ:VP} is straightforward.
\end{IEEEproof}

\begin{figure}
        \begin{center}
        \includegraphics[width=\figwidth]{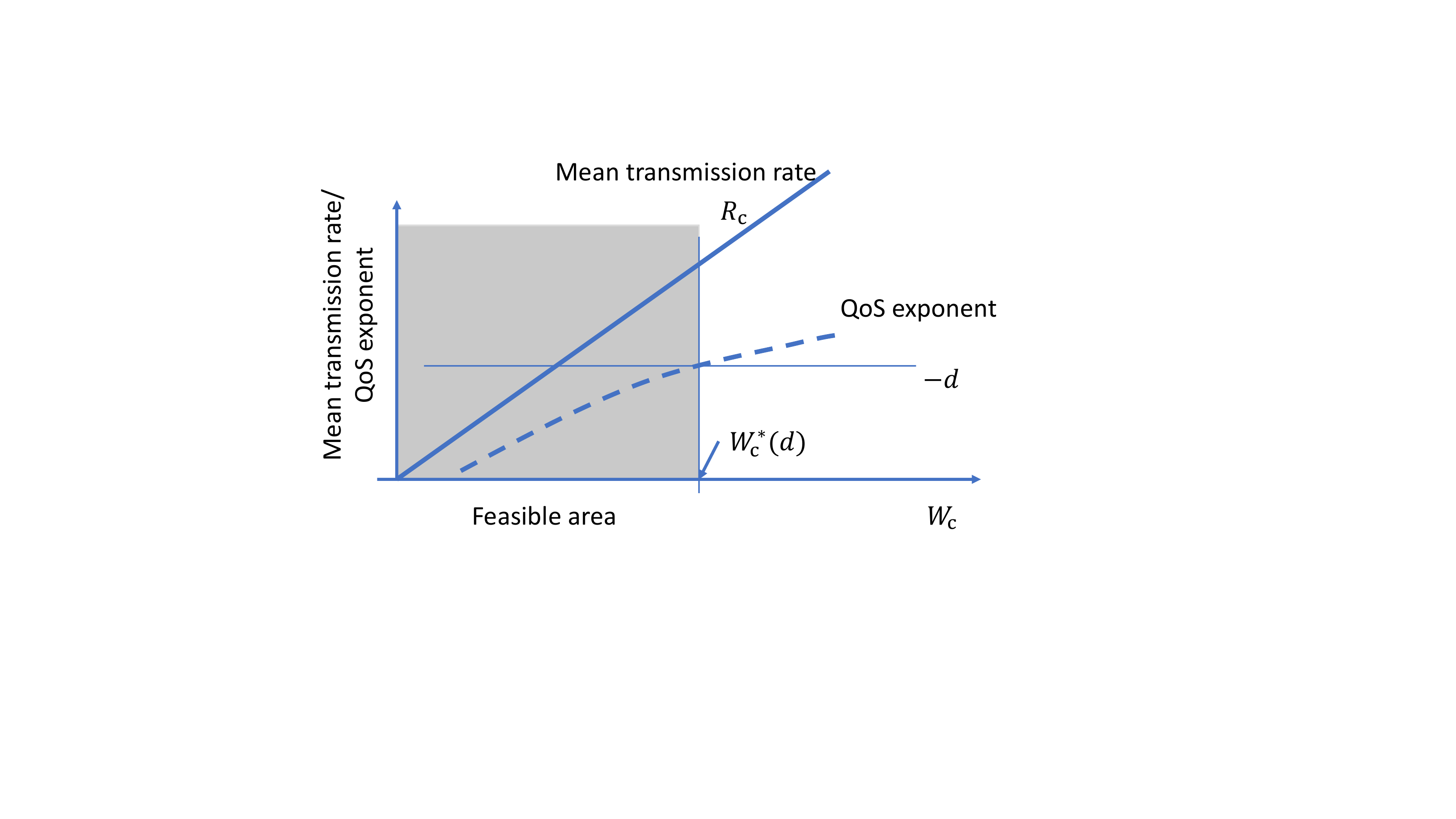}
        \end{center}
        \caption{An illustration of 
        $R_{\rm c}$ and $\psi_{\rm c} ( \theta (W_{\rm c}), W_{\rm c})$
        in the conventional approach.}
                \label{Fig:Fig3}
\end{figure}

For convenience, let $\lambda_0 = \lambda e^{-\frac{\lambda}{L}}$
and $\lambda_1 = \lambda e^{-\frac{\lambda}{L}} \bar s$.
Then, from Lemma~\ref{L:4}, it can be shown that
\begin{align}
M_y (W \theta) & = \lambda_0 ( e^{W \theta} - 1) \cr
M_v (\bar W \theta) & = \lambda_1 ( e^{\bar W \theta} - 1).
\end{align}
From this, $\psi(\cdot)$ can be re-written as
\be
\psi(\theta, W, \bar W) = 
\lambda_0 e^{W \theta} +
\lambda_1 e^{\bar W \theta} -(\lambda_0 + \lambda_1 + N \theta).
	\label{EQ:psi2}
\ee
Let $\theta (W, \bar W)$ be the solution 
that minimizes $\psi(\theta, W, \bar W)$.
From \eqref{EQ:psi2}, it can be shown that
$\theta (W, \bar W)$ is the solution of the following
equation:
\be
\lambda_0 W e^{W \theta} + \lambda_1 \bar W e^{\bar W \theta} = N.
	\label{EQ:ll}
\ee
Since the LHS in \eqref{EQ:ll} is an increasing function of $\theta$,
a sufficient condition for the existence of 
$\theta (W, \bar W) \ge 0$, which is unique, is 
\be
\lambda_0 W + \lambda_1 \bar W \le N.
\ee

Prior to discussing
the optimization of $W$ and $\bar W$
in EBT,
consider the optimization 
for the conventional 
approach where $W = \bar W = W_{\rm c}$.
In this case, it can be shown that
\be
\psi(\theta, W, \bar W) = 
\psi_{\rm c} (\theta, W_{\rm c})
\deft \lambda_{\rm c}(e^{W_{\rm c} \theta} - 1) - N \theta,
\ee
where $\lambda_{\rm c} = \lambda_0 + \lambda_1$.
It is trivial to show that $\theta$ minimizing
$\psi_{\rm c} (\theta, W_{\rm c})$ is given by
\be
\theta (W_{\rm c}) = \frac{1}{W_{\rm c}}
\ln \frac{N}{W_{\rm c} \lambda_{\rm c}} > 0, \ 
\mbox{if} \ N > W_{\rm c} \lambda_{\rm c}.
\ee
Thus, the minimum of $\psi_{\rm c} (\theta, W_{\rm c})$ becomes
\begin{align}
\min_{\theta \ge 0} \psi_{\rm c}(\theta; W_{\rm c}) 
& = \psi_{\rm c} ( \theta (W_{\rm c}), W_{\rm c}) \cr
& = 
\frac{N}{W_{\rm c}} \left(
1 - \ln \frac{N}{W_{\rm c} \lambda_{\rm c}} \right) - \lambda_{\rm c},
	\label{EQ:o_psi_c}
\end{align}
which is an increasing function of $W_{\rm c}$
(its derivative is positive when $N > W_{\rm c} \lambda_{\rm c}$).
Thus, for given QoS exponent $d$, 
the following rate maximization problem 
can be considered in the conventional approach:
\begin{align}
&\max_{W_{\rm c}} R_{\rm c} (W_{\rm c}) \cr
& \mbox{subject to} \ 
\psi_{\rm c} ( \theta (W_{\rm c}), W_{\rm c}) \le - d,
	\label{EQ:opt_c}
\end{align}
Since $R_{\rm c}$ increases with $W_{\rm c}$ as
in \eqref{EQ:R_c} and 
$\psi_{\rm c} ( \theta (W_{\rm c}), W_{\rm c})$ is also
an increasing function of $W_{\rm c}$, 
the solution becomes
\be
W_{\rm c}^* (d)
= \max
\left\{ W_{\rm c}: \psi_{\rm c} \left(
\theta(W_{\rm c}),W_{\rm c} \right) \le -d,
W_{\rm c} \in \uZ^+ \right\},
\ee
where $\uZ^+ = \{1, 2,\ldots \}$ represents
the set of positive integers.
In Fig.~\ref{Fig:Fig3}, we illustrate
$R_{\rm c}$ and 
$\psi_{\rm c} ( \theta (W_{\rm c}), W_{\rm c})$,
where $W_{\rm c}^* (d)$ is shown to be the solution.
Note that
due to the fact that $\min_{\theta \ge 0} 
\psi_{\rm c} (\theta, W_{\rm c})$
increases with $W_{\rm c}$, by letting $W_{\rm c} = 1$,
we can find the following condition 
for the existence of a feasible solution: 
\be
N \left(1 - \ln \frac{N}{\lambda_{\rm c}}\right) -
\lambda_{\rm c} \le -d, \ N \ge \lambda_{\rm c}.
	\label{EQ:feasible}
\ee

From Fig.~\ref{Fig:Fig3}, we can also observe
a trade-off relationship between the mean transmission
rate, $R_{\rm c}$, and the QoS exponent, $-d$.
That is, since the mean transmission rate decreases
when $-d$ decreases for a lower outage probability, 
it can be seen that the increase of 
the mean transmission rate results in a higher
outage probability, vice versa.

In EBT with the QoS constraint, the rate maximization problem
can be formulated as
\begin{align}
	\label{EQ:opt}
& \max_{W, \bar W} R(W, \bar W) \\
&\mbox{subject to} \
\left\{
\begin{array}{l}
\bar W \le W \cr
\min_{\theta \ge 0} \psi(\theta; W, \bar W) \le - d. \cr
\end{array}
\right.
\nonumber
\end{align}
Clearly, if $W = \bar W = W_{\rm c}$,
the problem in \eqref{EQ:opt}
reduces to that in \eqref{EQ:opt_c}.

\begin{figure}
        \begin{center}
        \includegraphics[width=\figwidth]{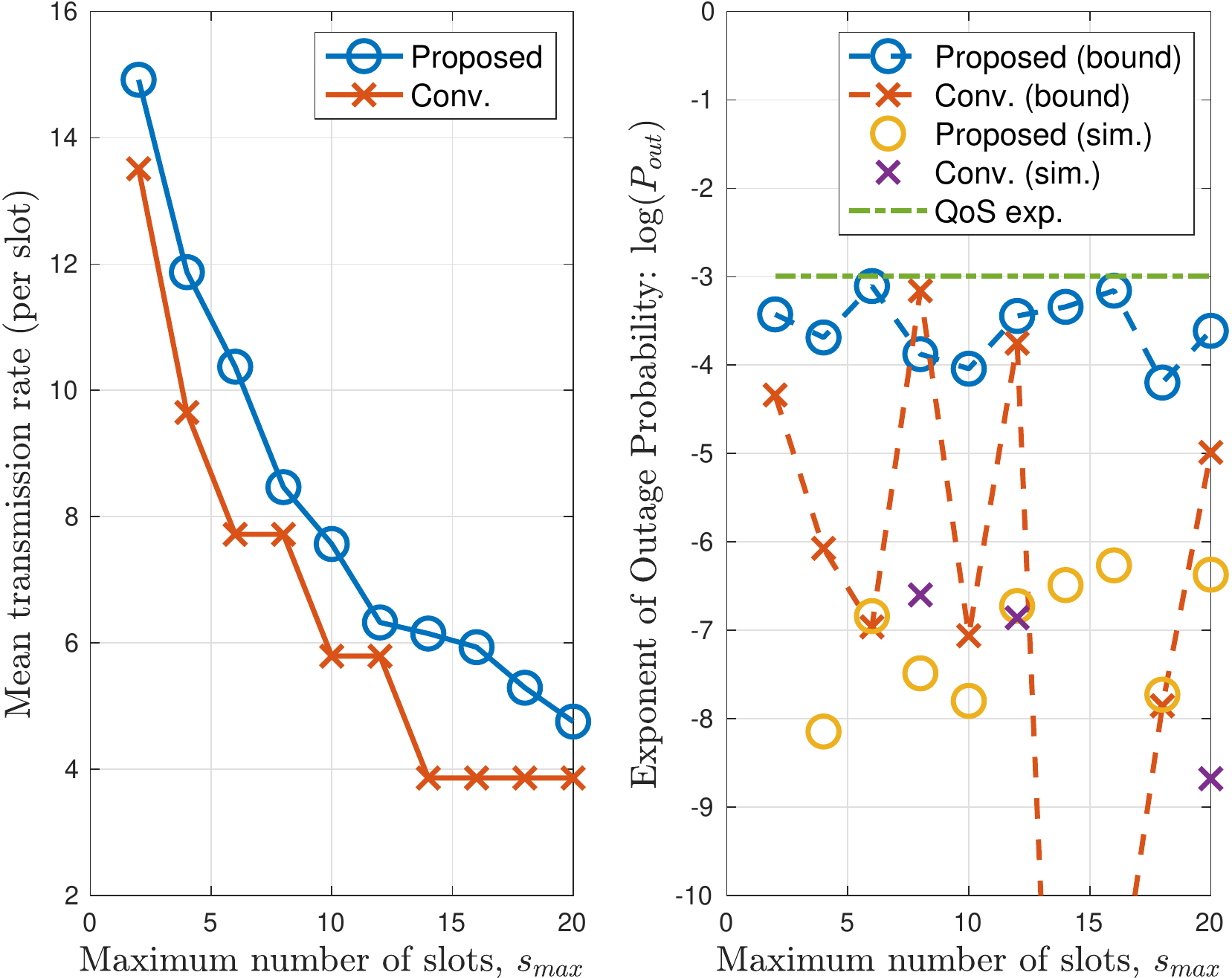} \\
        \hskip 0.8cm (a) \hskip 3.5cm (b) \\
        \end{center}
        \caption{Performance for different maximum number of slots
        for packet transmissions, $s_{\rm max}$,
        when $N = 200$, $\lambda = 8$, $L = 32$, and
        $d = -\ln(0.05)$ (i.e., $\Pr(Q > N) = 0.05$):
        (a) Mean transmission rate (per slot); (b) QoS exponent.}
                \label{Fig:plt1} \vskip -10.5pt
        \end{figure}
        
\begin{table}
        
        \caption{The numbers of channels for different 
        maximum number of slots
        for packet transmissions, $s_{\rm max}$,
        associated with Fig.~\ref{Fig:plt1}.}
        
        \centering
\resizebox{\tabwidth}{!}{\begin{tabular}{c|cccccccccc} \toprule
        $s_{\rm max}$& 2 & 4  &  6 &   8  & 10  & 12  & 14  & 16  & 18  & 20 \cr \hline
        $W$ &     8   & 7  &  7 &   7  &  5  &  8  &  7  &  6  &  4  &  3 \cr 
        $\bar W$& 7   & 5  &  4 &   3  &  3  &  2  &  2  &  2  &  2  &  2 \cr
        $W_{\rm c}$& 7& 5  &  4 &   4  &  3  &  3  &  2  &  2  &  2  &  2 \cr
        \bottomrule
        \end{tabular}}
        \label{TBL:1}
\end{table}

To find the optimal solution of the problem in
\eqref{EQ:opt}, we need to have few properties as follows.

\begin{myproposition}	\label{L:5}
For given $W$,
consider the following sub-problem of \eqref{EQ:opt}:
\begin{align}
& \max_{1\le \bar W \le W} R(W, \bar W)  \cr
& \mbox{subject to} \ \psi(\theta (W, \bar W); W, \bar W) \le - d,
	\label{EQ:L2}
\end{align}
where 
\be
\theta(W, \bar W) = \argmin_{\theta \ge 0} \psi(\theta; W, \bar W).
	\label{EQ:theta_WW}
\ee
Let
\be
\bar W^* (d, W) = \max\{\bar W :
\psi(\theta(W, \bar W); W, \bar W) \le -d, \bar W \in \uZ^+ \},
	\label{EQ:p_d}
\ee
which is, if exists, the optimal solution of \eqref{EQ:L2}.
\end{myproposition}
\begin{IEEEproof}
According to Lemma~\ref{L:3},
$\psi(\theta(W, \bar W); W, \bar W)$ is 
an increasing function of $\bar W$
(due to the strict inequality when $V$ is Poisson).
In addition, according to \eqref{EQ:RWW}, $R(W, \bar W)$ is
an increasing function of $\bar W$ for given $W$.
As a result, for given $W$, 
if \eqref{EQ:p_d} has a solution for a given $d$, it is unique
and the solution of \eqref{EQ:L2}.
\end{IEEEproof}

Thanks to the fact that 
$\psi(\theta(W, \bar W); W, \bar W)$ is 
an increasing function of $\bar W$,
for given $W$, we can find 
$\bar W^* (d, W)$ in \eqref{EQ:p_d} by incrementing
$\bar W$ by 1 from $\bar W = 1$.
While we can search for all $W$ from $W = 1$ to $N$
to find feasible $\{W, \bar W\}$,
the following property helps reduce the search interval.

\begin{figure}
        \begin{center}
        \includegraphics[width=\figwidth]{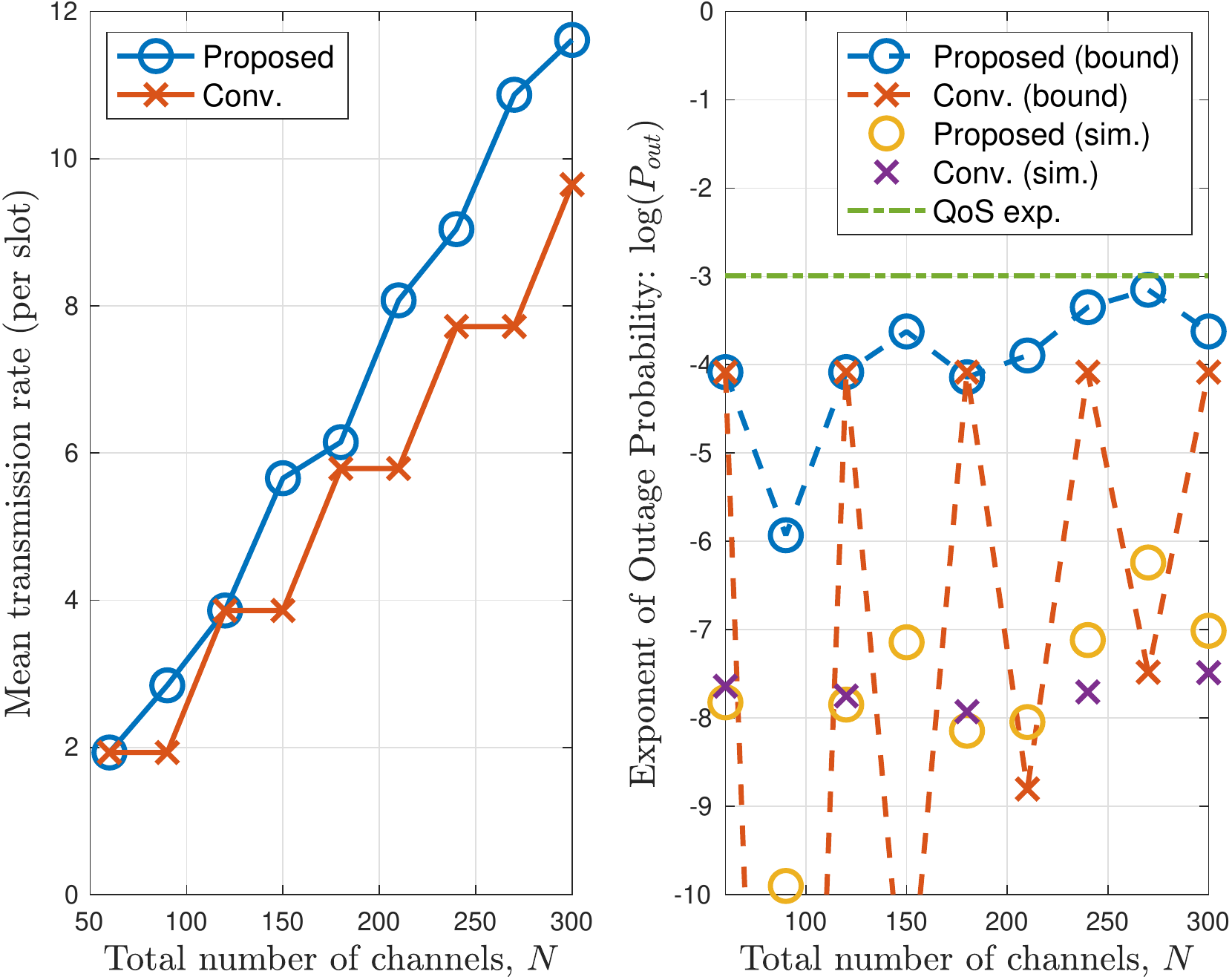} \\
        \hskip 0.8cm (a) \hskip 3.5cm (b) \\
        \end{center}
        \caption{Performance for different
        total number of channels, $N$,
        when $s_{\rm max} = 10$, $\lambda = 8$, $L = 32$, and
        $d = -\ln(0.05)$ (i.e., $\Pr(Q > N) = 0.05$):
        (a) Mean transmission rate (per slot); (b) QoS exponent.}
                \label{Fig:plt2}
        \end{figure}
        
        \begin{table}
        \caption{The numbers of channels for different
        total number of channels, $N$, 
        associated with Fig.~\ref{Fig:plt2}.}
        \centering
        \resizebox{\tabwidth}{!}{\begin{tabular}{c|cccccccccc} \toprule
        $N$ &  60 & 90  & 120 & 150 &  180 & 210 & 240 & 270 &  300 \cr \hline
        $W$ &     1   & 3  &  2 &   5  &  7  &  6  &  9  &  8  &  10  \cr
        $\bar W$& 1   & 1  &  2 &   2  &  2  &  3  &  3  &  4  &  4   \cr
        $W_{\rm c}$& 1& 1  &  2 &   2  &  3  &  3  &  4  &  4  &  5   \cr
        \bottomrule
        \end{tabular}}
        \label{TBL:2}
        \end{table}

\begin{mylemma}	\label{L:6}
Suppose that $\bar W^*(d, W)$ does not exist for given $W = W^\prime$.
Then, it is also true for 
any $W^{\prime \prime}$ that is greater than~$W^\prime$.
\end{mylemma}

Since it can be easily proved by using the fact
that $\psi(d, W, \bar W)$ is also an increasing function of $W$,
we do not provide details for the proof of Lemma~\ref{L:6}.
Note that 
$\min_{\theta \ge 0} \psi(\theta; W, \bar W)$ becomes the smallest
if $W = \bar W = 1$. Thus, in order to find any $W \ge 1, \bar W \ge 1$,
the following condition is required:
\be
\min_{\theta \ge 0} \psi(\theta; 1, 1) \le - d,
\ee
which is equivalent to \eqref{EQ:feasible}.

Based on the above results,
in order to obtain a set of feasible pairs $\{W, \bar W\}$,
we can have the algorithm in 
Algorithm~\ref{AL:1}.
Once the set of feasible $\{W, \bar W\}$ is
obtained, the optimal $\{W, \bar W\}$
is the pair that has the highest $R(W, \bar W)$.

Note that the complexity of
Algorithm~\ref{AL:1}
depends on 
the complexity of finding
$\bar W(d,W)$ in \eqref{EQ:p_d}.
Since $\bar W \le W$, for each value of $W$,
$\bar W \in \{1, \ldots, W\}$. Suppose that 
for a given pair of $(\bar W, W)$,
the complexity to find $\theta(W,\bar W)$ in \eqref{EQ:theta_WW} is
$\Omega_0$, which is independent of $W$ and $\bar W$. 
Then, the total complexity
is at most $\Omega_0 \sum_{W = 1}^N W 
= \Omega_0 \frac{N(1+N)}{2}$.
Furthermore, since 
$\psi(\theta; W, \bar W)$ is a convex function of 
$\theta$, $\Omega_0$ depends on a numerical technique for 
1-dimensional convex optimization.
However, in general, $\Omega_0$ is inversely proportional
to a required precision.

\begin{algorithm}[t]
        \SetAlgoLined
        \KwResult{Optimal $\{W, \bar W\}$}
         Inputs: $\{\lambda_0, \lambda_1, N, d\}$\;
         Initialization: $W = 1$, $\cR = \{ \}$, $\bar \cW = \{\}$\;
         \While{ $W \le N$}{
          find $\bar W^*(d, W)$ (using Lemma~\ref{L:5})\;
          \eIf{$\bar W^*(d, W) \ge 1$}{
           $\bar \cW \Leftarrow \bar \cW \cup \bar \{W^*(d, W)\}$\;
           $\cR \Leftarrow \cR \cup \bar \{R(W, \bar W^*(d, W))\}$\;
           $W \Leftarrow W + 1$\;
           }{
           Stop (thanks to Lemma~\ref{L:6})\;
          }
         }
         Output: $\{W, \bar W\}$ corresponding to the highest $R$ in~$\cR$\;
        
         \caption{Finding optimal $\{W,\bar W\}$}
                \label{AL:1}
\end{algorithm}

In summary, to compare the overall performance of 
the conventional and proposed approaches,
we need to find the mean rates after optimizing
$(W, \bar W)$ for the proposed approach via \eqref{EQ:opt}
and $W_{\rm c}$ for the conventional approach via \eqref{EQ:opt_c}
for given key parameters, e.g., $N,\lambda, L$, and $d$.
Unfortunately, it is difficult to find closed-form expressions
for the maximum mean transmission rates.
Thus, we need to rely on numerical methods to find them,
which will be considered in Section~\ref{S:Sim}.

\begin{figure}
        \begin{center}
        \includegraphics[width=\figwidth]{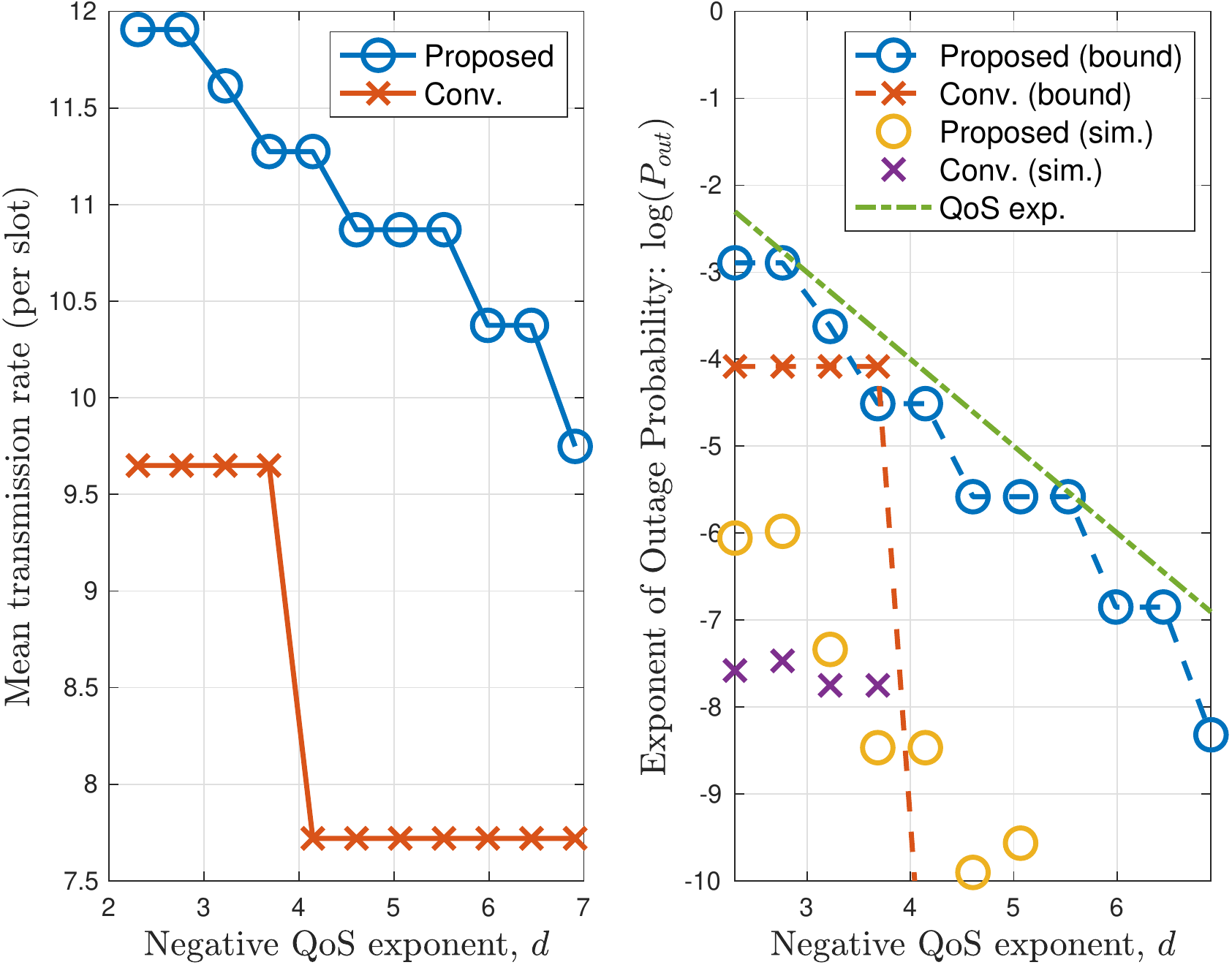} \\
        \hskip 0.8cm (a) \hskip 3.5cm (b) \\
        \end{center}
        \caption{Performance 
        for different negative QoS exponent, $d$,
        when $N = 300$, $s_{\rm max} = 10$, $\lambda = 8$, and $L = 32$:
        (a) Mean transmission rate (per slot); (b) QoS exponent.}
                \label{Fig:plt3} \vskip -7.5pt
\end{figure}
        
        \begin{table}
                \caption{The numbers of channels for different 
                negative QoS exponents, $-d$,
                associated with Fig.~\ref{Fig:plt3}.}
                \centering
        \resizebox{\tabwidth}{!}{\begin{tabular}{c|ccccccccccc}\toprule
                $\jh{d}$ &  2.3 & 2.7  & 3.2 & 3.6 &4.1&4.6&5.0&5.5&5.9&6.4 &6.9\cr \hline
                $W$ &     11  &11  & 10 &   9  &  9  &  8  &  8  & 8 &  7  &  7  &  6 \cr 
                $\bar W$& 4   & 4  &  4 &   4  &  4  &  4  &  4  & 4 &  4  &  4  &  4 \cr
                $W_{\rm c}$& 5& 5  &  5 &   5  &  4  &  4  &  4  & 4 &  4  &  4  &  4 \cr
        \bottomrule        
        \end{tabular}}
                \label{TBL:3}
                \end{table}

\section{Simulation Results}	\label{S:Sim}

In this section, we present simulation results
to see the performance of the proposed and
conventional approaches 
with $\{W,\bar W\}$ and $W_{\rm c}$
by solving \eqref{EQ:opt_c} and \eqref{EQ:opt}
under the assumption
of independent Rayleigh fading channels in \eqref{EQ:Ray}.
For convenience, we assume that $\bar \gamma = 6$ dB for all UEs.
Furthermore, independent Poisson arrivals are assumed for the UEs
that send randomly selected preambles 
from a pool of $L$ preambles
with mean arrival
rate $\lambda$.
For the number of slots for packet transmissions, $s$,
is assumed to have a uniform distribution so that
$\nu_s = {1}/{s_{\rm max}}$, $s = 1, \ldots, s_{\rm max}$.

In Fig.~\ref{Fig:plt1},
we show the performance for different maximum numbers of slots
for packet transmissions, $s_{\rm max}$,
in terms of the mean transmission rate and QoS exponent
when $N = 200$, $\lambda = 8$, $L = 32$, and
$d = -\ln(0.05)$ (i.e., $\Pr(Q > N) \le 0.05$).
In addition, the corresponding 
optimal numbers of channels for exploration 
and exploitation (i.e., $W$ and $\bar W$, respectively)
are shown together with $W_{\rm c}$ in Table~\ref{TBL:1}.
It is shown that 
the mean transmission rate per slot, 
$R(W,\bar W)$, of EBT 
is higher than that of the conventional approach, $R_{\rm c} (W_{\rm c})$,
for a wide range of $s_{\rm max}$,
while all the QoS exponents 
are less than $-d \approx 2.9957$ (or $\Pr(Q > N) = 0.05$).
We also see that the number of channels per UE (i.e.,
$W$ and $\bar W$ in EBT and $W_{\rm c}$ in the conventional approach) decreases with $s_{\rm max}$ 
in Table~\ref{TBL:1}, which also results in the decrease
of the mean transmission rate as shown in Fig.~\ref{Fig:plt1}.
This is expected as there are more packets
to be transmitted from UEs with a larger $s_{\rm max}$.

\begin{figure}
        \begin{center}
        \includegraphics[width=\figwidth]{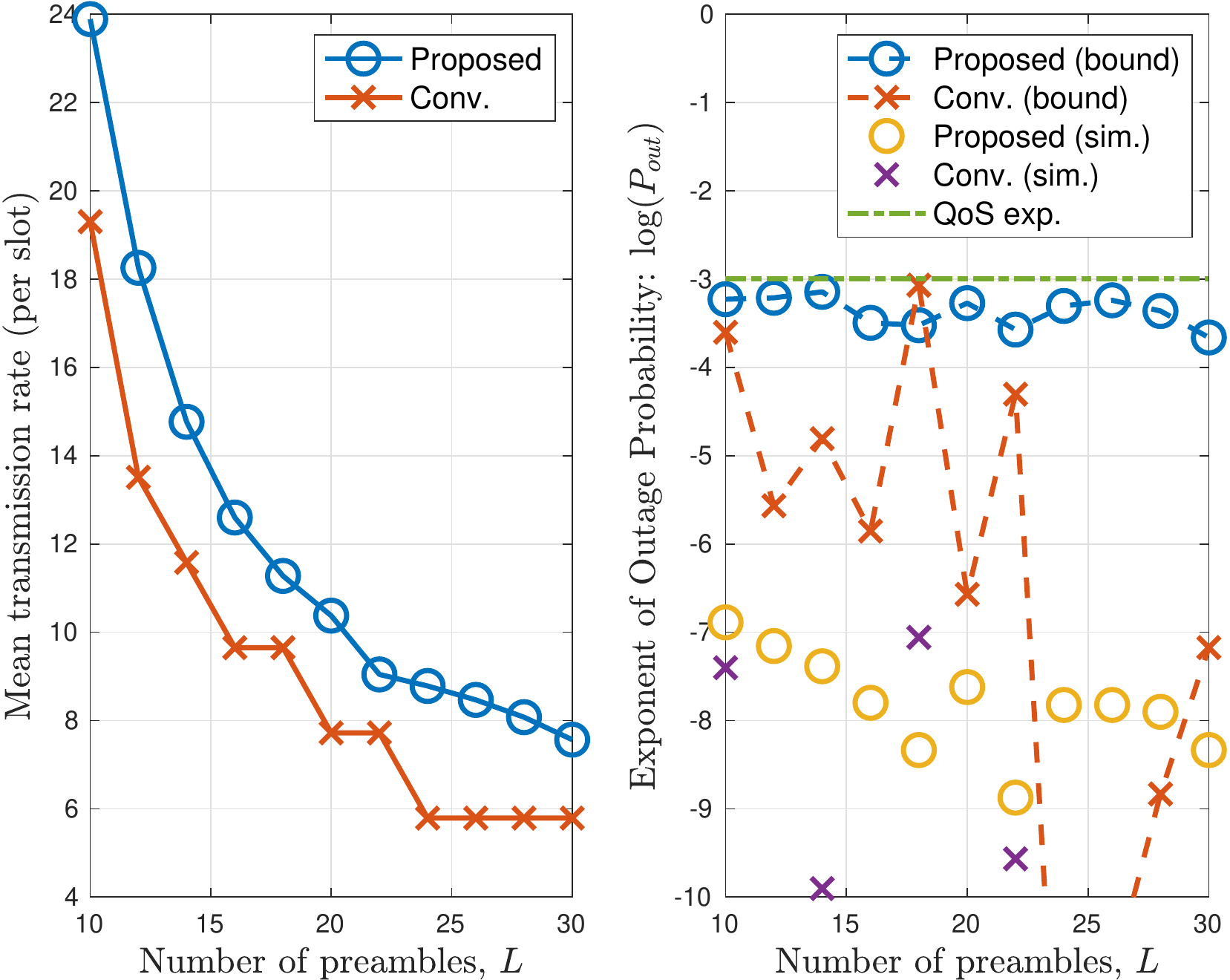} \\
        \hskip 0.8cm (a) \hskip 3.5cm (b) \\
        \end{center}
        \caption{Performance for different
        numbers of preambles, $L$,
        when $N = 300$, $s_{\rm max} = 10$, $\lambda = 20$, and 
        $d = -\ln(0.05)$ (i.e., $\Pr(Q > N) = 0.05$):
        (a) Mean transmission rate (per slot); (b) QoS exponent.}
                \label{Fig:plt5} \vskip -7.5pt
        \end{figure}
        
        \begin{table} \label{TBL:5}
        \caption{The numbers of channels for different
        numbers of preambles, $L$, associated with Fig.~\ref{Fig:plt5}.} \vskip -3pt
        \centering
        \resizebox{\tabwidth}{!}{\begin{tabular}{c|ccccccccccc}\toprule
        $L$ &  10& 12 & 14 &  16  & 18  & 20  & 22  & 24  & 26  &  28 & 30 \cr \hline
        $W$ &    16   &12  &  9 &   8  &  9  &  7  &  9  &  8  &  7  &  6 & 5 \cr
        $\bar W$& 9   & 7  &  6 &   5  &  4  &  4  &  3  &  3  &  3  &  3 & 3\cr
        $W_{\rm c}$&10& 7  &  6 &   5  &  5  &  4  &  4  &  3  &  3  &  3 & 3 \cr
        \bottomrule
        \end{tabular}}
        \end{table}

The impact of the total number of channels, $N$,
on the performance is illustrated in 
Fig.~\ref{Fig:plt2}
when $s_{\rm max} = 10$, $\lambda = 8$, $L = 32$, and
$d = -\ln(0.05)$ (i.e., $\Pr(Q > N) = 0.05$).
As shown in 
Fig.~\ref{Fig:plt2},
the increase of $N$ results in the increase
of the mean transmission rate with keeping the QoS exponent lower
than $-d$. Clearly, the increase 
of the mean transmission rate can be achieved by
increasing $W$ and $\bar W$ in EBT
and $W_{\rm c}$ in the conventional approach as shown in
Table~\ref{TBL:2}.
It is interesting to see a particular case that $N = 300$.
According to Table~\ref{TBL:2},
the optimal numbers of channels for exploration and exploitation
in EBT are $W = 10$ and $\bar W = 4$,
respectively, while $W_{\rm c}$ in the conventional approach
is $5$. 
Since $s_{\rm max} = 10$, we have $\bar s = 5.5$.
On average, each UE in the conventional approach
sends a total of $(\bar s +1) W_{\rm c} = 6.5 \times 5 = 
32.5$ packets\footnote{It includes pilot signals transmitted
by a UE to allow the BS to estimate the channel coefficients
and SNR.}. 
On the other hand,
in EBT, each UE sends
a total of $W + \bar s \bar W = 10 + 5.5 \times 4 = 32$ packets 
on average. That is, the numbers of transmitted packets
in both the approaches are almost the same. In this case,
while their QoS exponents are comparable,
we can see that the mean transmission of EBT 
is higher 
than that of the conventional approach as shown in Fig.~\ref{Fig:plt2}.

As discussed earlier,
the optimal numbers of channels can be found
by solving \eqref{EQ:opt_c} and \eqref{EQ:opt}
for a given QoS exponent.
In Fig.~\ref{Fig:plt3},
we show the performance
for different negative QoS exponent, $d$,
when $N = 300$, $s_{\rm max} = 10$, $\lambda = 8$, and $L = 32$.
As shown in Fig.~\ref{Fig:plt3} (b),
it is possible to keep the upper-bound on the 
QoS exponent below $-d$ by choosing
optimal $W$ and $\bar W$ in EBT.
We see that the mean transmission rate of EBT is higher than that of the conventional approach
for any value of $d$.

In Table~\ref{TBL:3},
we can see the optimal numbers of channels for each UE
associated with Fig.~\ref{Fig:plt3}.
There are two variables, $W$ and $\bar W$,
for the optimization in EBT, 
while there is only one variable, $W_{\rm c}$,
in the conventional approach. Consequently,
it is possible to jointly
optimize the pair of $W$ and $\bar W$ 
so that the resulting QoS exponent is sufficiently close to $-d$
in EBT. However, in the conventional approach,
the optimization result is coarse as there is only one variable,
i.e., $W_{\rm c} \in \{1, 2,\ldots\}$, to be optimized, which results in 
a QoS exponent that is a bit loose.
For example, when $d = - \ln (0.01) \approx 4.6$,
the optimal number of channels in the conventional approach
is $W_{\rm c} = 4$.
However, with $W_{\rm c} = 5$, it results in $d = 4.083$,
which is slightly lower than the target negative QoS exponent, $d = 4.6$.
From this, we can see that another salient advantage
of EBT over the conventional one
is that the radio RBs can be more 
precisely allocated to meet a QoS requirement
thanks to an additional variable in terms of optimization perspective.

In Fig.~\ref{Fig:plt5},
we show the performance for different
numbers of preambles, $L$,
when $N = 300$, $s_{\rm max} = 10$, $\lambda = 20$, and 
$d = -\ln(0.05)$.
Since $\lambda$ is fixed, as $L$ increases,
there are less preamble collisions,
which leads to more admitted UEs per slot in the third step.
As a result, we can see that the mean
transmission rate decreases with $L$,
while the QoS exponents are all less than $-d$ due to
optimal allocation of channels per UE (as shown in
Fig.~\ref{Fig:plt5} (b)).
It is also observed that the numbers of channels
per UE, $W$ and $\bar W$, in EBT decrease with $L$,
which is also true in the conventional approach
as shown in Table~\ref{TBL:3}.

According to Figs.~\ref{Fig:plt1} (b) - \ref{Fig:plt5} (b),
it has been observed that 
the actual QoS exponents obtained by simulations are lower than
the QoS exponents that are obtained
from the theory (e.g., \eqref{EQ:o_psi_c} for
the conventional approach).
The gap is due to a number of reasons as follows:
\emph{i)} the Poisson approximation in \eqref{EQ:P_Z} and
\eqref{EQ:Y_s};
\emph{ii)} the Chernoff bound in \eqref{EQ:CB}.
In particular, since $Z$ in \eqref{EQ:Z_B}
is a binomial variable, $Z$ cannot be larger than $L$.
However, in \eqref{EQ:P_Z}, 
due to the Poisson approximation, 
$Z$ can be larger than $L$, which results in 
a heavier tail probability of $Q$ than the actual one
and a loose upper-bound in \eqref{EQ:CB}.
Thus, we need to consider different methods 
to find tighter bounds as a further research topic.
Other further research topics are to be discussed in
Section~\ref{S:Con}.

\section{Concluding Remarks}	\label{S:Con}

In this paper, we proposed EBT that can increase the mean transmission rate
by exploiting the multichannel selection diversity gain 
for uplink transmissions in MTC.
To exploit the channel selection diversity gain,
in particular,
a BS initially allocated $W$ channels to a new admitted UE
in the third step and estimated their SNRs. Then, 
the BS sent the indices of the best $\bar W$ channels
so that the UE can use them to upload their data packets,
where $\bar W < W$.
In general, for a fixed $\bar W$, it was shown that a higher 
multichannel selection
diversity gain can be achieved as $W$ increases.
Compared with the conventional approach
where $W = \bar W = W_{\rm c}$, it was shown that
EBT can have an improved transmission rate.
However, since the use of more channels for exploration
can result in a higher outage probability,
an optimization problem was formulated
with an outage probability constraint 
to allow a fair comparison with 
the conventional approach
from a multiuser system point of view.
We also derived an algorithm to find the optimal 
numbers of channels for exploration and exploitation,
i.e., $W$ and $\bar W$, in EBT.
From simulation results, we confirmed that
EBT can provide a higher mean
transmission rate than the conventional one under various conditions
thanks to multichannel selection diversity.

There are a number of issues to be studied in the future.
Firstly, time-varying channels need to be studied,
while we only focus on static channels in this paper.
In particular, for mobile UEs, it is necessary to generalize
the channel selection to take into account time-varying channels.
Second, a more realistic objective function
(instead of mean transmission rate) 
can be considered with finite-length codes and
re-transmission protocols (e.g., hybrid automatic
repeat request (HARQ) protocols \cite{WickerBook} \cite{LinBook}).
Third, as mentioned earlier, it is interesting to find
a tighter upper-bound on the outage probability
as the Chernoff bound with Poisson approximation turns to be a bit loose.

\bibliographystyle{ieeetr}
\bibliography{mtc}

\end{document}